\documentstyle[prb,aps,epsfig,multicol]{revtex}
\begin{document}
\sloppy
\draft
\title{Thermodynamic properties of Holstein polarons and the effects 
of disorder}
\author{ A. N. Das $^{a}$ and S. Sil $^{b}$}
\address {$^{a}$   Theoretical Condensed Matter Physics Division,
 Saha Institute of Nuclear Physics, \\
1/AF, Bidhannagar, Kolkata 700064, India\\
$^{b}$ Department of Physics, Visva Bharati, Santiniketan- 731 235,
India \\}
%\date{\today}
\maketitle
%%%%%%%%%%%%%%%%%%%%%%%%%%%%%%%%%%%%%%%%%%%%%%%%%%%%%%%%%%%%%%%%%%%%%
\begin{abstract}
%\maketitle
The ground state and finite temperature properties of polarons
are studied considering a two-site and a four-site Holstein model
by exact diagonalization of the Hamiltonian. The kinetic energy,
Drude weight, correlation functions involving charge and lattice 
deformations, and the specific heat have been evaluated as a function
of electron-phonon ({\it e-ph}) coupling strength and temperature. 
The effects of site diagonal disorder on the above properties have 
been investigated. The disorder is found to suppress the kinetic 
energy and the Drude weight, reduces the spatial extension of the 
polaron, and makes the large-to-small polaron crossover smoother. 
Increasing temperature also plays similar role.
For strong coupling the kinetic energy arises mainly from the 
incoherent hopping processes owing to the motion of electrons within the 
polaron and is almost independent of the disorder strength.
From the coherent and incoherent contributions to the 
kinetic energy, the temperature above which the incoherent part
dominates is determined as a function of {\it e-ph} coupling strength.
\end{abstract}
%\pacs{~71.38., 63.20.kr}
%\maketitle
%==============================================================================
\begin{multicols}{2}
\begin{center}
{\bf I. Introduction}
\end{center}
\vskip 0.3cm
Study of different properties of polarons has been of great importance
since the evidence of polaronic charge carriers in many materials of 
recent interest, viz. high-$T_c$ cuprates \cite{hightc}, CMR-manganites
 \cite{zhao}, biological materials like DNA \cite{DNA}, etc. which have 
large technological potential and importance. 
In the simplest Holstein model an electron in a narrow
tight-binding band interacts locally with dispersionless optical
phonons. For large {\it e-ph} coupling the resultant
polaron is a small polaron with high effective mass, while
for weak coupling it becomes a large polaron having a
much lower effective mass for a finite adiabatic parameter.
The crossover from a large to a small polaron and the corresponding
changes in the polaronic properties in the ground state have been
studied for the Holstein model by different groups \cite 
{DS,dasprn,prndas,Jeck,Fehske98,Capone,Romero,Trugman,jcdas,Barisic,zolidas} 
using various methods to enrich our understanding in this field. 
However, finite temperature 
study of the properties of polarons and the effect of disorder 
on polaronic properties are few and needs more attention.
Previously we have studied  the effect of disorder on some of 
the polaronic ground state properties {\it (i)} for a 
two-site system following a perturbation method based on a modified 
Lang-Firsov (MLF) phonon basis \cite {jdas} and
{\it (ii)} for a many-site system following a zero-phonon averaging of 
the MLF-transformed hamiltonian and a real space renormalization group 
method to deal with the disorder \cite{dassil}. 
The above studies have the limitations that they are only for the ground 
state. The first study, though quite accurate, has been carried out only 
for a two-site system,
while the latter study gives approximate results. 
In this paper we will consider a two-site and a four-site Holstein
model and follow an exact diagonalization method to study the ground
state as well as finite temperature properties of the polarons and the
effect of disorder on them. We will mainly study the kinetic energy,
correlation functions involving charge and lattice deformations, 
Drude wight and the specific heat of the systems
as a function of {\it e-ph} coupling for different temperatures and 
disorder strength. 
 
The paper is organized as follows. In Section II we have developed the 
formalism for the aforementioned  study considering the Holstein
model. We have presented the results and discussions for the 
two-site Holstein model in Section III-A and those for the four-site
system in Section III-B. The conclusion is given in Section IV. 
In Appendix-A we have shown analytically considering an infinite size 
system and following strong-coupling second order perturbation theory that 
the disorder has weak or negligible effect on the kinetic energy 
for strong coupling.      

\begin{center}
{\bf II. Formalism }
\end{center}
\vskip 0.3cm
The Holstein Hamiltonian with site diagonal disorder in 1-d is given by
\begin{eqnarray}
H &=&\sum_{i} \epsilon_i c_i^{\dag} c_{i}
     -  \sum_{i} (t c_i^{\dag} c_{i+1} + h.c)
    +  g \omega  \sum_i n_i (b_i^{\dag} + b_i) \nonumber\\
       &+ & \omega \sum_i b_i^{\dag} b_i
\end{eqnarray}
$c_i^{\dag}$ and $c_i$ are the electron creation and annihilation 
operators at the site $i$, $n_i~( = c_i^{\dag} c_i)$ is the
number operator, $b_i^{\dag}$ and $b_i$ are the creation
and annihilation operators for the phonons corresponding
to interatomic vibrations at site $i$ and $\omega$ is the phonon frequency.
Electronic hopping takes place only between the nearest-neighbor sites  
with hopping strength t and g denotes the local $e$-ph coupling. 
The electronic site energy $\epsilon_i$ is independent of the site $i$ 
for the ordered case. To study the effect of the site-diagonal disorder
we would put a different site potential at one of the sites of the 2- 
or 4-site system.
Spin index is not used for the electron, because a single polaron case 
has been studied here. 
                                 
The third and fourth terms of Eq.1 represent the electron-phonon 
interaction and phonon harmonic energy, respectively. These terms 
may be written in the momentum space defined by the phonon creation 
operators: 
$b_{\bf q}^{\dagger}$= $(1/{\sqrt N})~\sum_i b_i^{\dagger} e^{i \bf{q.R_i}}$ 
and the corresponding annihilation operators, where $N$ is the number of
sites in the system \cite {dasprn,prndas,zolidas}.
It can be easily shown that the in-phase (${\bf q}=0$) phonon mode 
does not couple with the electron dynamics but with the total number 
of electrons of the system. The harmonic term of this phonon mode 
along with its interaction 
with the electron may be separated out and written in a diagonal form 
\cite{dasprn,prndas}.
The rest of the Hamiltonian involving $(N-1)$ phonon modes and
$N$ electronic states (for a single electron problem) are considered
to construct the eigen basis and matrix elements for diagonalization 
of the matrix. If one considers $n_p$
number of phonon states per mode then the total number of basis
states will be $n_{Tot}=N n_p^{(N-1)}$. Elimination of the in-phase 
mode, thus, reduces the states of the Hilbert space by a factor of 
$n_p$, which is an advantage for any diagonalization procedure.

For the electron states we use the site space basis, which is 
convenient to take into account of the site disorder. For the phonon
states we use the momentum space basis so that the in-phase (${\bf q=0}$)
mode may be separated out.
The Hamiltonian is then diagonalized to obtain the eigenstates and 
the eigenenergies. Thermodynamic expectation value of any observable
characterized by the operator $O$ is then found out by
\begin{eqnarray} 
\langle O \rangle &=& \frac{1}{Z} \sum_{n=1}^{n_{Tot}} \langle n|O|n \rangle 
e^{-\beta E_n} \\
 Z &=& \sum_{n=1}^{n_{Tot}} e^{-\beta E_n}     
\end{eqnarray}      
where $E_n$ is the eigen energy of the $n$-th eigenstate $|n\rangle$, 
$\beta = 1/k_BT $ and $T$ denotes the temperature.
In this paper we are interested in evaluating the kinetic energy, static
correlation function involving charge and lattice deformation,
specific heat and the Drude weight. 
The operator corresponding to the kinetic energy is
\begin{equation}
H_t=  - t \sum_{i, j}^{\prime} c_i^{\dag} c_{j}.
\end{equation}
For the correlation functions involving charge and lattice 
deformations we calculate
\begin{equation}
\chi_m(i)= \langle n_i (b_{i+m}^{\dagger} + b_{i+m}) \rangle/2g
\end{equation}
which represents the lattice deformation produced at the site $i+m$ 
when the electron is at site $i$. The Drude weight ($D_n$) in units 
of  $\pi e^2$ for an eigenstate $|n\rangle$ of
the polaronic system is obtained by introducing a phase factor to the 
hopping matrix element ($t \rightarrow t e^{i \phi}$) in order to break 
the time reversal symmetry  
and then finding out the response of the break down of the time reversal 
symmetry to the electric current as \cite{kohn}
\begin{equation}
D_n =  \frac{\partial^2E_n(\phi)}{\partial \phi^2} |_{\phi =0}
\end{equation} 
where $E_n(\phi)$ is the eigen energy of the $n$-th eigenstate in
presence of non zero $\phi$.
The thermodynamic expectation value of the Drude weight ($D$) 
is found out by taking the thermal average of $D_n$ over all the eigen 
states 
\begin{equation}
\langle D \rangle = \frac{1}{Z} \sum_n D_n e^{-\beta E_n}
\end{equation}
The specific heat may be expressed in terms of the energy fluctuation of the 
system at a finite temperature $T$ as
\begin{equation}
C_v/k_B= \frac{1}{(k_B T)^2}~ [\langle E^2 \rangle - \langle E \rangle^2]  
\end{equation}  
where $\langle E \rangle$ and $\langle E^2 \rangle$ are the thermal 
average of the energy and the square of the energy respectively.

\begin{center}
{\bf III. Results and discussions}
\end{center}
\vskip 0.3cm
{\bf A. Two-site system :} \\
\vskip 0.2cm
For the 2-site system the electron dynamics is coupled only to the 
out-of-phase ($\bf q=\pi$) phonon mode. The Hamiltonian, which has 
to be considered for numerical diagonalization, is \cite{jdas}
\begin{eqnarray}
H_d &=&  \sum_{i} \epsilon_i n_{i} 
- t (c_{1}^{\dag} c_{2} + c_{2}^{\dag} c_{1}) \nonumber \\
&+& \omega  g_{+} (n_1-n_2) (d + d^{\dag}) +  \omega  d^{\dag}d 
\end{eqnarray} 
where $g_{+}=g/\sqrt 2$ and $d=~(b_1-b_2)/\sqrt 2 $.
We consider the basis states $c_{i}^{\dagger}|0 \rangle_e |n_d\rangle_{ph}$,  
where $i$=1,2 
and $n_d$= 0, 1, 2, ..., $n_d^{m}$, is the number of phonons in 
the d-oscillator.
We refer to this as the bare basis. The Lang-Firsov transformed or
MLF-transformed basis for the d-oscillators may also be used for 
diagonalization. But the matrix elements of the Hamiltonian operator
are much simpler in the bare basis than those in other (LF or MLF)
basis. We find that for exact diagonalization study the convergence 
of the results is achieved with much less number of phonon states 
within the bare 
basis compared to the LF or MLF basis, although the MLF basis is 
best and much better than the bare basis for perturbation 
calculation \cite{jcdas}. 

To show the convergence of the correlation functions in the bare basis 
we have evaluated 
$\langle n_1u_2 \rangle$, where $u_2= (b_2+b_2^{\dagger})$, using 
different values of $n_p$, and plotted it as a function of $g_+$ for 
different values of $t$ and $\epsilon_d~ (=\epsilon_2 - \epsilon_1)$ 
in Fig.1.  We will refer to $\epsilon_d$ for the two-site system as 
the disorder strength since it partly mimics the role of disorder 
in larger systems.
It is found that in the range $0 \le g_+ \le 3.2$ the results obtained 
for $n_p=25$ is quite accurate. The results for $n_p=$ 25, 30 and 40 are 
indistinguishable. 
We also find that for a particular value of $n_p$ the highest value 
of $g_+$, up to which the convergence is achieved, almost does not 
depend on $\epsilon_d$ or $t$ (Fig.1) in the range of parameters 
we have studied.

In Fig. 2 we have plotted the kinetic energy evaluated for the 
parameters $t=$1 and $\epsilon_d$=1, using different values of $n_p$.
It is found that the curves are indistinguishable for $n_p \ge 15$   
in the range $0< g_+ < 3.2$. 
It shows that a much smaller value of $n_p$ is sufficient to achieve 
the convergence in the kinetic energy than that for the convergence 
in the correlation function $\langle n_1(b_2+b_2^{\dagger})\rangle$.
In the following we will present the results for the two-site system 
for $n_p$=40 or 50. The latter is used for high temperature and strong
coupling. 
 
In Fig. 3 we have presented the variation of the correlation function 
$\chi_m(i) = \langle n_i u_{i+m} \rangle /2g $ for $i$=1,2 and $m$=0,1 
for the parameters $t=2.1$ and $\epsilon_d = 1$. 
In the absence of disorder $\chi_m(i)$ is independent of $i$.
An introduction of the disorder ($\epsilon_d \ne 0$) breaks the 
translational symmetry of the system and $\chi_0(1)$ and $\chi_0(2)$ 
become different such that $\chi_0(1)- \chi_0(2)$ increases with the 
increase of the strength of the disorder potential ($\epsilon_d$). 
However, it is observed that $\langle n_1u_2 \rangle 
= \langle n_2u_1 \rangle$ even in the presence of disorder. 
For a periodic Holstein model  
the on-site correlation involving charge and lattice deformation 
$\langle n_i u_i \rangle $ is always larger than the intersite correlation
$\langle n_i u_{i \pm 1} \rangle $. 
With increasing {\it e-ph} coupling $\langle n_i u_i \rangle $ increases 
while $\langle n_i u_{i \pm 1} \rangle $ decreases indicating a crossover 
from a large polaron to a small polaron.
In Fig.3 it is seen that in presence of disorder the correlation 
$\langle n_2 u_2 \rangle$ 
decreases with increasing {\it e-ph} coupling and in the strong coupling 
region $\langle n_2 u_2 \rangle < \langle n_2 u_1 \rangle $. These are 
very different from the normal behavior of polarons in ordered systems. 
However, $\langle n_1 u_1 \rangle $ and $\langle n_1 u_2 \rangle $ 
show the usual characteristics of the polaron crossover with increasing 
{\it e-ph} coupling, 
The above-mentioned observations create confusion in distinguishing 
the large and small 
polaron from the structure of the correlation function $\chi_m(i)$ in the 
disordered polaronic system. In this situation we define average correlation 
functions
\begin{eqnarray}  
\chi_0^{avg} &=& \frac{\sum_{i} \chi_0(i)}{\sum_i n_i} \\
\chi_1^{avg} &=& \frac{\sum_{i} \chi_1(i)}{\sum_i n_i} \\
\chi_d^{avg} &=& \frac{\sum_{i}(\chi_0(i)- \chi_1(i))}{\sum_i n_i}
\end{eqnarray}  
to describe the polaronic behavior for the disordered polaronic system. 

In Fig. 4 we have shown the variations of the kinetic energy and 
$\chi_d^{avg}$ with $g_+$ for $t=2.1$ for different values of
$\epsilon_d$. It is seen that the kinetic energy is suppressed
with increasing {\it e-ph} coupling as well as with increasing disorder 
strength. In the intermediate range of coupling the kinetic energy 
shows an exponential suppression. For strong coupling
the kinetic energy shows a $1/g^2$ behavior and is almost
independent of the disorder strength, as noted previously \cite {jdas}. 
In the Appendix-A, considering an infinite lattice Holstein model 
we have shown analyticallly that for strong coupling the polaronic
hopping is so small that the kinetic energy arises mainly from 
incoherent hopping owing to the undirected motion of the electron 
keeping the centre of the polaron fixed. This incoherent hopping 
contribution for large $g$ is almost independent of the disorder strength
and inversely proportional to  $g^2$.
Fig. 4 shows that with increasing disorder strength 
the correlation $\chi_d^{avg}$ increases, which indicates that the size 
of the polaron becomes smaller, and the crossover (from a large to
a small polaron) occurs at a lower value of $g_+$.
For strong coupling, disorder has almost no effect on $\chi_d^{avg}$ as
similar to that observed for the kinetic energy. It may be noted that
increasing disorder strength makes the polaron crossover more smoother. 

To examine the effect of temperature on the properties of polarons 
we have evaluated the kinetic energy and the correlation function  
for different temperatures and disorder strength. In Figs. 5 and 6 
we have plotted $\chi_d^{avg}$ and the kinetic energy for $t=2.1$ 
as a function of $g_+$ for different temperatures. For weak and 
intermediate coupling $\chi_d^{avg}$ increases,  
implying that the size of the polaron becomes smaller, while
the kinetic energy is suppressed  with increasing temperature. 
For stronger coupling ($g_+ > 1.7$) the temperature has an opposite effect,
{\it i.e.}, the kinetic energy increases with temperature. However,
the effect of temperature is small for strong coupling.    
At high temperature the variation of the kinetic energy with {\it e-ph}
coupling is much weaker compared to that at low temperatures.
It is also found that disorder has very 
little effect on both the kinetic energy and the correlation function 
at high temperatures (not shown in the figure).

We have studied the specific heat ($C_v$) of the ordered as well as 
disordered polaronic system for different temperatures and hopping. 
In Fig. 7. we have presented the $C_v$ of the ordered two-site system as a 
function of {\it e-ph} coupling for different temperatures for the 
hopping parameter $t$=2.1. Similar plots for different disorder 
strengths ($\epsilon_d= 0, 0.5$ and $1.0$), and for $t$=1 and 2.1 are shown 
in Fig.8.

Fig.7 shows that in the low temperature regime the specific heat 
shows a peak at intermediate coupling. With increasing temperature 
the peak shifts towards a lower value of $g_+$ and then disappears 
while a dip 
is developed in the intermediate coupling region. At low temperatures 
the specific heat is mainly governed by the separation of the ground state 
and the first excited state of the two-site Holstein model.

It may be noted that the specific heat for a system having only two energy 
levels with energy separation ($\Delta E$), shows a peak at 
$\Delta E/ k_B T= 2.58$, and $C_v$ is  very small when $\Delta E$ is 
far from $2.58 k_BT$. 
For the two-site Holstein model the energy separation ($\Delta E$) between two 
lowest eigen energy levels decreases monotonically with $g_+$ and becomes 
negligibly small at strong coupling \cite {ranninger}. 
%It may be noted that the specific heat for a system having only two energy 
%levels with energy separation ($\Delta E$), shows a peak at 
%$\Delta E/ k_B T= 2.58$, and $C_v$ is  very small when $\Delta E$ is 
%far from $2.58 k_BT$. 
At very low temperatures $C_v$ is very small for weak {\it e-ph} coupling, 
as $\Delta E/k_BT$ is sufficiently large. $C_v$ attains a maxima at an 
intermediate coupling, 
when $\Delta E$ is $\sim 2.58 k_BT$, and becomes very small 
in the strong coupling region as $\Delta E \rightarrow 0$. 
At a higher temperature a higher value of $\Delta E$ is 
required to achieve the maximum value in $C_v$. Since $\Delta E$
increases with decreasing $g_+$ for the two-site Holstein model, 
the peak in $C_v$ is obtained at a lower value of $g_+$ for a higher
temperature. 
With increasing temperature the higher energy states, in addition to the 
ground and first excited states, have significant contributions to the 
specific heat. This leads to the absence of the peak and formation of a 
dip in $C_v$.   
In Fig. 8 a comparison of the variation of specific heat as a function 
of $g_+$ for the hopping parameters $t = 1$ and $2.1$ is given. 
In absence of any disorder the energy separation $\Delta E$ increases 
with the increase of the hopping parameter t, hence a shift in the 
position of the peak in $C_v$ is obtained at a higher value of $g_+$.

The effect of disorder on the specific heat of a polaron is also shown 
in Fig. 8. In presence of disorder the specific heat is suppressed 
for weak and 
intermediate coupling. However, for strong coupling the $C_v$ 
is larger compared to that for the ordered case.

\vskip 0.5cm
{\bf B. Four-site system :} \\
\vskip 0.2cm

For the four site Holstein model, out of the four phonon modes there 
are three modes with ${\bf q} \ne 0$ which couple to the electron 
dynamics. 
The four-site Holstein model Hamiltonian may be divided into two
parts: one part containing the electronic terms, harmonic terms
of the three phonon modes and the interaction of these phonon modes
with the electron number operators, while the second part gives 
a diagonal form for the shifted in-phase (${\bf q}=0$) oscillator 
(see Eqs. 7 and 8 of Ref.6). The first part, which contains the 
non trivial physics of the system and cannot be treated exactly
by any analytical method, is diagonalized numerically. 
We have done the numerical diagonalization with 
$n_p$=9 per phonon mode and checked that this gives fairly accurate 
results in the range of our study presented here. 
For the disordered 
case we break the translational symmetry by introducing a different 
site-potential at one of the lattice
sites while keeping the equal site-potential for the rest.        

For the Holstein model the correlation functions satisfy a sum 
rule 
\begin {eqnarray}
\sum_m \chi_m(i) &=& \sum_m \langle n_iu_{i+m} \rangle /2g \nonumber \\  
                  &=& \langle n_i \rangle
\end {eqnarray}
In absence of disorder $\langle n_iu_j \rangle = \langle n_ju_i \rangle$
because of translational symmetry and a relation 
$\sum_i \langle n_iu_j \rangle /2g = \langle n_j \rangle $ 
is satisfied. We find that even for the disordered Holstein model
$\langle n_iu_j \rangle = \langle n_ju_i \rangle$, 
though the sites $i$ and $j$ have different site potentials, and 
the relation 
$\sum_i \langle n_iu_j \rangle /2g = \langle n_j \rangle $ 
is also satisfied for the disordered system in addition to the 
sum rule (13). 
As mentioned previously average correlation functions should be
used for the disordered case to characterize the nature (large or small) 
of the polaron. Averaging also justifies that the disordered impurity  
may occupy any site at random. The average correlation 
functions involving charge and lattice deformations are:
\begin{eqnarray}  
\chi_m^{avg} &=& \frac{\sum_{i} \chi_m(i)}{\sum_i n_i} 
\end{eqnarray}  
where $m=0, 1, 2$. 

In Fig. 9 we have plotted the kinetic energy (scaled to 2$t$)
and $\chi_0^{avg}$ against g for the ordered case and for 
disordered cases with site potentials (1,0,0,0) and (-1,0,0,0),
where the values of the site potentials ($\epsilon_i$) for 
$i=1,2,3,4$ are shown within the parentheses (all the energies
are expressed in units of $\omega=1$). It is seen that
the polaron becomes more localized (in size) with higher value of 
$\chi_0^{avg}$ and lower kinetic energy for disordered cases compared to the 
ordered case. For the (-1,0,0,0) case the value of $\chi_0^{avg}$
is much higher and the kinetic energy is much lower than those for the 
(1,0,0,0) case, because in the former case the electron will tend to
be trapped at the site of the negative potential. This would suppress
the kinetic energy and favor small polaron formation in presence of
{\it e-ph} coupling. In the same figure we have also shown the variations of
$\chi_1$  and $\chi_2$ for the ordered case. $\chi_1$  and $\chi_2$ 
represent the lattice deformations produced at the nearest- and  
next-nearest neighbor sites of an electron. 
For small $g$ the values of $\chi_1$  and $\chi_2$ are appreciable 
indicating that the polaron has spread over the lattice. With increasing $g$,
$\chi_1$  and $\chi_2$ reduce and become very small for strong coupling,
the corresponding polaron is a small one. For the disordered cases
the $\chi_1^{avg}$  and $\chi_2^{avg}$ behave in the same way (not
shown in the figure), but their values are lower than those for 
the ordered case and become insignificant at a lower value
of $g$ compared to the ordered case. In the strong coupling limit 
the dependence of the kinetic energy and the correlation function
on the disorder strength is very weak. The reasons have been explained 
in Appendix-A.

We have studied the effect of temperature on the correlation functions
$\chi_0$, $\chi_1$ and $\chi_2$ and on the kinetic energy.       
Fig.10 shows the variation of the on-site correlation function ($\chi_0$) 
and the kinetic energy ($-\langle K \rangle$) with temperature for 
different $g$ values. Except for strong coupling the $\chi_0$ 
increases while the kinetic energy 
decreases with increasing temperature. It may be noted that when 
$\chi_0$ increases, the values of $\chi_1$ and $\chi_2$ decrease
and the polaron size becomes smaller.
For strong coupling the effect of temperature is small but opposite.
The above behavior point to the fact that the polaron gets more and 
more localized with increasing temperature in the regime of weak and 
intermediate {\it e-ph} coupling, while for very strong coupling 
the polaron size gets larger with increasing temperature.
We have also studied the properties of the ground state and
different excited states individually to get a clear understanding
of the observed temperature variation.
It is found that in the regime of weak and intermediate coupling 
the value of $\chi_0$ is larger while the kinetic energy is smaller 
for the excited states compared to those of the ground state and 
this leads to increase (decrease) of $\chi_0$ ($-\langle K \rangle$) 
with increasing temperature. 
We have given a representative plot showing the variation of $\chi_0$ 
with energy for the ground and the excited states for $g=1$ in Fig. 11.
For strong coupling ($g \ge 2.5$) we find that the $\chi_0$ is 
smaller for the excited states (in general) than that for the ground 
state and the polaron becomes a larger polaron with increasing temperature. 
 
For a fixed temperature  $\chi_0$ increases and the kinetic energy 
decreases with increasing $g$ (Fig. 10), but their variation with $g$ 
is much slower at higher temperatures.
These results are qualitatively similar to that noted for the 
two-site system. It may be mentioned that Hohenadler \cite {Hohen} 
studied the 1-d Holstein model by quantum Monte Carlo method and observed 
similar variation of the kinetic energy with increasing temperature.

In Fig. 12 we have shown the variation of $C_v$ as a function of $g$
for the ordered and disordered systems for $T=0.1$.     
The specific heat shows a peak at intermediate coupling, which 
is suppressed with increasing disorder strength. The behavior of $C_v$ for 
the site potentials (-0.5,0,0,0) is similar to that observed for the 2-site
case with $\epsilon_d=0.5$. In Fig. 12 we have also shown the plots 
of $\chi_0^{avg}$ to show that the specific heat peak occurs in the 
region of $g$ where $\chi_0^{avg}$ also undergoes a sharp change. 
However, the peak position in $C_v$, as mentioned previously, would be 
mainly decided by the tuning of the energy separation of two lowest 
levels of the system with the thermal energy at low temperatures. 

The kinetic energy contains contributions from both the coherent and
incoherent hopping processes \cite {Fehske}.
The Drude weight (in units of $\pi e^2$)
represents the coherent part of the kinetic energy. The contribution
from the incoherent hopping processes may be found out from the
total kinetic energy and the Drude
weight by using the f-sum rule (see appendix B),
\begin{eqnarray}
- \frac{\langle K \rangle}{2} &=& \frac{\langle D \rangle}{2} +
  \langle S^{reg} \rangle
\end{eqnarray}
where, $\langle K \rangle $ and $\langle D \rangle$ are the thermal
average of the kinetic energy and the Drude weight, respectively, and
\begin{equation}
S^{reg}  = t^2  \sum_{n' \ne n}\frac{|\langle n | \sum_i (c_i^\dagger
c_{i+1} - c_{i+1}^\dagger c_i)| n' \rangle|^2}{E_{n} - E_{n'}}
\end{equation}
is proportional to the contribution to the kinetic energy from the
incoherent hopping processes. 
%$\langle \psi_n^0 |$ and $E_{n} $ are
%the $n$th eigen state and eigen energy of the Hamiltonian concerned.
 
In Fig. 13 we have shown the effect of disorder on the Drude weight
of the four-site Holstein model for $t=0.5$.  
The kinetic energy and the Drude weight are plotted against $g$ for 
the ordered and disordered systems.  
At $g$=0, the Drude wight and the kinetic energy
have same values for the ordered case indicating that in absence
of {\it e-ph} interaction and disorder the entire part of the kinetic 
energy comes from coherent hopping. 
In the range of intermediate to strong
coupling the Drude weight shows an exponential suppression.
The kinetic energy, on the other hand,  shows an exponential suppression
only in the range of intermediate coupling and a $1/g^2$ behavior in
the strong coupling region, as predicted by the strong coupling 
perturbation theory. For the disordered case the Drude weight    
is smaller than the kinetic energy even at $g$=0, because of 
disorder-induced incoherent hopping. 

We have investigated the effect of temperature on the coherent and 
incoherent parts of the kinetic energy by evaluating the kinetic
energy and the Drude weight as a function of temperature for different 
values of $g$. Some representative plots (for $g$= 0.1, 0.5, 1 and
1.5) are given in Fig.14. At low temperature the kinetic energy 
as well as the Drude weight show negligible dependence on temperature,
represented by the flat region of the curves. This flat region is larger 
for smaller $g$. After the flat region the kinetic energy reduces 
exponentially with temperature. At high temperatures, where the Drude
weight is very small, the kinetic energy may be fitted to a function 
$a/T+bT$. For small value of $g$ the value of $b$ is very small 
and the kinetic energy varies approximately as $1/T$. 
In the temperature range $1.5 \le k_BT \le 2.5$ the values of $a$ 
and $b$ are, respectively,  
0.477 and 0.003 for $g=0.1$, 0.450 and 0.0054 for $g=0.5$,
0.376 and 0.012 for $g=1.0$ and  0.281 and 0.020 for $g=1.5$.
The Drude weight shows almost exponential dependence on temperature
both in the intermediate and high temperature range. 
A good fit to the Drude weight is obtained with a function 
$a e^{-bT}$, but with different values of the parameters $a$ and $b$ 
in the two (intermediate and high temperature) ranges. 

The incoherent part of the kinetic energy can be directly determined 
from the difference between the negative kinetic energy and the Drude weight. 
In Fig. 14 we have also shown the variation of the incoherent part as a 
function of temperature for different $g$ values. 
Except for very strong coupling (not shown in the figure) the incoherent 
part increases with
temperature, reaches a peak and then decreases at a slow rate with 
increasing temperature. 
A cross-over temperature ($T_{cross}$) may be defined from the intersection 
of the curves for the coherent and the incoherent part of the kinetic
energy such that for $T< T_{cross}$ the coherent part is dominant and for
$T> T_{cross}$ the incoherent part is dominant. We have plotted 
the variation of $T_{cross}$ with $g$ in Fig. 15. The cross-over 
temperature decreases with $g$. The $T_{cross}$ {\it vs.} $g$ curve
shows a sudden change in the gradient at $g$=1 for $t=0.5$. 
The rate of fall of $T_{cross}$ with $g$ is higher 
in the region $g>1$ than that for $g<1$. 

To get the combined effect of the temperature and the disorder 
we have studied the Drude weight and incoherent part of the kinetic  
energy as a function of temperature for the site potentials (0,0,0,0)
and (-1,0,0,0). In Fig.16 we have given such a plot for 
intermediate coupling ($g=1$) where the effect of disorder
is large. It is seen that at low temperatures the effect of 
disorder is very large on both the Drude weight and the 
incoherent part of the kinetic energy. The Drude weight is
rapidly suppressed while the incoherent part is enhanced
a lot by the disorder site potential. 
At high temperatures where the
Drude weight becomes very small, the disorder has little 
effect on the incoherent part of the kinetic energy. The
effect of the disorder on the kinetic energy is even smaller
because in this range of temperature the Drude weight slightly
decreases while the incoherent part slightly increases with
the introduction of the disorder. 
   
We would address now the possible size effect on
the results. In the Appendix-A we have derived a strong coupling
effective polaronic hamiltonian where the different coefficients 
(in eq. A-2) have no size dependence provided the number of nearest 
neighbors ($z$) is same. For a four-site system, $z=2$ which is
same as that for an infinite 1-d chain. The kinetic energy in the 
ground state, obtained from Eq. (A2), for $z=2$ is given by
\begin{eqnarray}
K_{G} &=&  - 2 t_p - 4 \frac {t_p^2}{\omega}
\sum_{n,m =0; n+m \ge 1 }^\infty
\frac{g^{2(n+m)}}{n!~ m!~(n+m)}  \nonumber \\
&-& 4 \frac{t_p^2}{\omega} \sum_{n=1}^\infty \frac{g^{2m}}{m!~m}
\end{eqnarray}
where $t_p=t e^{-g^2}$. 
The above  clearly shows that for strong coupling the kinetic energy 
in the ground state should not have any size dependence if the number 
of nearest neighbors
is same, hence the kinetic energy for $N= 4$ will be same as that 
for an infinite lattice, where $N$ is the number of lattice sites. 
This is completely consistent with the results obtained in a recent numerical 
study by Hohenberg et al. \cite {Hohen}. They obtained the same kinetic 
energy for $N=4,8,16,32$ for large $g$ while a small size effect is
observed for small and intermediate coupling. With increasing temperature
the size effect is even smaller as noted in Ref.19.

\vskip 1.0cm
\begin{center}
{\bf IV. Conclusions}
\end{center}
\vskip 0.3cm

We have investigated the effect of disorder and temperature on the 
properties of a polaron for a two- and four-site Holstein model. It 
has been observed that both the disorder and temperature reduce the 
polaron size and suppress the kinetic energy in the weak and intermediate 
{\it e-ph} coupling regime. For strong {\it e-ph} coupling polarons 
are practically immobile and the kinetic energy arises mainly from
the to and fro motion of the (bare) electron between nearest-neighbor 
sites keeping the centre of the polaron fixed. In this regime the effects 
of disorder and temperature on the correlation functions and the kinetic 
energy are very small. 

The polaronic kinetic energy has contributions from the coherent and the 
incoherent hopping processes. The contribution from the coherent hopping 
decreases with increasing {\it e-ph} coupling, temperature and strength 
of the disorder. For small values of $g$ and $T$ this contribution 
(Drude weight) shows a weak suppression with $g$ or $T$ and then 
a rapid suppression with increasing $g$ (or $T$). 
The contribution from the incoherent hopping, on the other hand, increases
with increasing $g$ or $T$ initially, reaches a maximum and then decreases.
This contribution decreases as $1/g^2$ with $g$ in the strong coupling. 
In the regimes of strong coupling or high temperature
the coherent contribution to the kinetic energy is very small
compared to the incoherent contribution and the latter becomes almost 
independent of the disorder strength.      
We have also identified a cross-over 
temperature below which coherent part of the kinetic energy is dominant 
and above which the incoherent part is dominant. This cross over 
temperature decreases with the increase of the electron-phonon coupling. 

The variation of the polaronic specific heat with respect to the {\it e-ph}
coupling shows a peak in the intermediate {\it e-ph} coupling regime at
low temperatures. A suppression of the specific heat due to 
disorder has been observed when {\it e-ph} coupling is in the weak or  
intermediate regime. 

For the Holstein model the interaction is very short ranged 
and the size effect generally does not play a significant role
in shaping different properties \cite {Hohen,Capone2,Takada}.
As pointed out previously the size effect on the kinetic 
energy is negligible for strong coupling, while the effect 
is finite but small for weak and intermediate coupling. 
At higher temperatures the size effect is even smaller.
Regarding the role of the disorder on the Drude weight of
the polaronic system, the size effect has not been reported
to our knowledge. We find that for strong coupling or at higher 
temperatures the effect of disorder on the polaron dynamics 
is negligible, hence the size effect will also be negligible
as similar to that for the ordered system. 
In other region of the coupling the qualitative behavior 
of the results presented here for the four-site system 
are expected to be the same for larger systems.
\vskip 1.0cm

%\appendix
\renewcommand{\theequation}{A-\arabic{equation}}
\setcounter{equation}{0}
\section*{APPENDIX- A}
%\begin{center}
%{\bf Appendix-I}
%\end{center}
%\vskip 0.3cm
 Applying the standard Lang Firsov (LF) transformation to the Holstein model
in Eq. (1) one obtains 
\begin{eqnarray}
\bar{H} &=& e^R H e^{-R} \nonumber \\
  &=& \sum_{i} (\epsilon_i - \epsilon_p) c_i^{\dag} c_{i}
     -  {\sum_{i,j}}'  t_p c_i^{\dag} c_j e^{g(b_i^{\dag}-b_j^{\dag})}
e^{-g(b_i-b_j)} \nonumber \\
       &+ & \omega \sum_i b_i^{\dag} b_i 
\end{eqnarray}
 where $R=\sum_i g n_i(b_i-b_i^{\dag})$, $\epsilon_p= g^2 \omega$ is the 
polaron binding energy, $t_p=e^{-g^2}$ is the polaronic hopping strength 
and $j$ is a nearest neighbor of $i$.

The second term of Eq.(A-1) represents the kinetic energy operator which
involves zero-phonon as well as multiphonon processes (in the LF
phonon basis) associated with
the hopping of a polaron between nearest-neighbor sites. In the 
strong coupling limit, where $t_p$ is very small, following a second order 
strong coupling perturbation theory an effective expression 
for the kinetic terms (within the phonon ground state) may be obtained as
\cite {Romero2}
\begin{eqnarray}
H_{kin} &=&  - t_p {\sum_{i,j}}' c_i^{\dag} c_j  \nonumber \\  
&-& 2zt_p^2 
\sum_{n,m =0; n+m \ge 1 }^\infty
\frac{g^{2(n+m)}}{n!~ m!~(n+m) \omega} \sum_i c_i^\dag c_i \nonumber \\
&-& 2 t_p^2 \sum_{n=1}^\infty \frac{g^{2m}}{m!~m \omega}
{\sum_{i,k}}^" c_i^\dag c_k 
\end{eqnarray}
where $z$ is the number of nearest neighbors.
The first term of  $H_{kin}$ represents the coherent hopping of
polarons without emission or absorption of phonons, the second term 
originates from virtual hopping of a polaron from site $i$ to a 
nearest neighbor site $j$ and back. This hopping is associated with 
virtual emission of multi phonons at the sites $i$ and $j$ followed
by absorption of all the emitted phonons while hopping back. In this 
process the (bare) electron within the polaron undergoes forward 
and backward motion between nearest-neighbor sites
keeping the polaron immobile.  
Third term represents an effective hopping of a polaron from site $i$ 
to a second nearest neighbor site $k$ via a common nearest neighbor 
site $j$ where multi phonons are created and then absorbed.
For strong coupling the coefficients of the first and third terms 
in Eq. (A-2) become negligible and only the second term 
($\sim t^2/\epsilon_p$) contributes to the kinetic energy. 
In presence of disorder $(n+m)\omega$ in the denominator 
of the second term modifies to $(n+m)\omega \pm \epsilon_d$, where    
$\epsilon_d$ is the difference between the site potentials at sites 
$i$ and $j$.
The major contributions of this term for strong coupling comes from 
multi phonon processes with high values of $(n+m) \omega$ such that
$(n+m) \omega >> \epsilon_d$. As a consequence of the above reasons the 
coefficient of the second term in (A2) as well as the kinetic 
energy in the strong coupling region becomes very weakly dependent on 
the disorder potential.

It may be mentioned that the Eq.(A2) is valid for any dimension 
and size of the system. The value of $z$ would be different
for different dimensions. 
\vskip 1.0cm
\vskip 0.3cm
\renewcommand{\theequation}{B-\arabic{equation}}
\setcounter{equation}{0}
\section*{APPENDIX- B}
 The Hamiltonian of the system in the broken time reversal symmetry
may be written as
\begin{equation}
H(\phi) = -\sum_i (t e^{i \phi} c_i^\dagger c_{i+1} + h.c) + V
\end{equation}
where $V$ is composed of electronic potential and interaction of electrons 
with lattice. Let $H_0$ be the Hamiltonian of the system when $\phi =0$, 
and the corresponding eigen energies and eigen states be $E_n^0$ 
and $| \psi_n^{0} \rangle$, respectively. 
If $E_n(\phi)$ be the eigen energy of the Hamiltonian $H(\phi)$, 
the Drude weight for 
the $n$-th eigen state of the Hamiltonian $H_0$ is \cite{kohn}
\begin{equation}
D_n = \frac{\partial^2 E_n(\phi)}{\partial \phi^2~~~~} |_{\phi=0}
\end{equation}

For small $\phi$, $H(\phi)$ can be expressed as
\begin{eqnarray}
H(\phi) &=& H_0 -i \phi t \sum_i(c_i^\dagger c_{i+1} -c_{i+1}^\dagger c_i)
\nonumber \\ 
&+& \frac{\phi^2}{2} t
\sum_i(c_i^\dagger c_{i+1} + c_{i+1}^\dagger c_i ) + ...... 
\end{eqnarray}
Considering the terms involving $\phi$ as a perturbation a second order
perturbation calculation gives 
\begin{eqnarray}
E_n(\phi) &=& E_n^0 - i \phi t\langle \psi_n^0 | (\sum_i c_i^\dagger c_{i+1}
- c_{i+1}^\dagger c_i)| \psi_n^0 \rangle \nonumber \\
&+& \phi^2 t^2 \sum_{n' \ne n}\frac{|\langle \psi_n^0 | (\sum_i c_i^\dagger 
c_{i+1} - c_{i+1}^\dagger c_i)| \psi_{n'}^0 \rangle|^2}{E_{n'}^0 - E_n^0} \nonumber \\
&+& \frac{\phi^2 t}{2} \langle \psi_n^0 |\sum_i (c_i^\dagger c_{i+1} + 
c_{i+1}^\dagger c_i ) |\psi_n^0 \rangle + .... 
\end{eqnarray}
Therefore 
\begin{eqnarray}
D_n &=& 2 t^2 \sum_{n'\ne n}\frac{|\langle \psi_n^0 | (\sum_i c_i^\dagger 
c_{i+1} - c_{i+1}^\dagger c_i)| \psi_{n'}^0 \rangle|^2}{E_{n'}^0 - E_n^0}
\nonumber \\
& - & K_n 
\end{eqnarray}
where,  
\begin{equation}
K_n = -t \langle \psi_n^0 |\sum_i (c_i^\dagger c_{i+1} + c_{i+1}^\dagger c_i ) 
|\psi_n^0 \rangle  
\end{equation}
is the expectation value of the kinetic energy for the $n$-th 
eigen states of the Hamiltonian $H_0$. Taking the thermal average over 
all the eigenstates at a temperature $T$ establish the sum rule 
\begin{eqnarray}
&-& \frac{\langle K \rangle}{2} = \frac{\langle D\rangle}{2} \nonumber \\
&+& t^2 \langle \sum_{n' \ne n}\frac{|\langle 
\psi_n^0 | \sum_i (c_i^\dagger 
c_{i+1} - c_{i+1}^\dagger c_i)| \psi_{n'}^0 \rangle|^2}{E_{n} - E_{n'}} \rangle 
\end{eqnarray}
where the thermal average of an observable $A$ is  
\begin{equation}
\langle A \rangle = \frac{\sum_n A_n exp(-E_n/k_BT)}{\sum_n exp(-E_n/k_B T)}.
\end{equation}
In the above expression $A_n$ is the expectation value of the observable 
$A$ for the $n$-th eigen state of the Hamiltonian $H_0$. 
%\end{appendix}
\vskip 0.3cm

%\begin{thebibliography}{}

\end{multicols}

\newpage

\begin{figure}
\resizebox*{3.1in}{2.5in}{\rotatebox{270}{\includegraphics{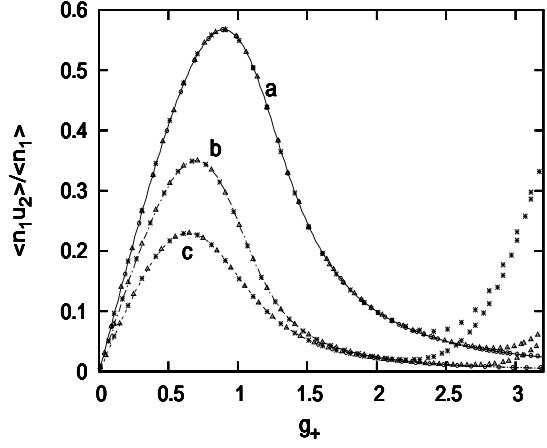}}}
\vspace*{0.5cm}
\caption[]{
Convergence of the correlation function $\langle n_1u_2 \rangle /
\langle n_1 \rangle$ 
($u_2=b_2+b_2^{\dagger}$) for the two-site system. (a) Solid line: 
$t$= 2.1, $\epsilon_d$= $\epsilon_2-\epsilon_1$=1.0 with $n_p$=40; 
(b) dashed line : $t$=1.0, $\epsilon_d$= 0.5 with $n_p$=40;
(c) dash-dotted line : $t$=1.0, $\epsilon_d$=1.0 with $n_p$=40.
Solid squares are for $n_p$=15, solid triangles are for $n_p$=20,  
solid circles are for $n_p$=25.  
All energy parameters are expressed in units of $\omega$=1.
}
\label{scaling}
\end{figure}
\begin{figure}
\resizebox*{3.1in}{2.5in}{\rotatebox{270}{\includegraphics{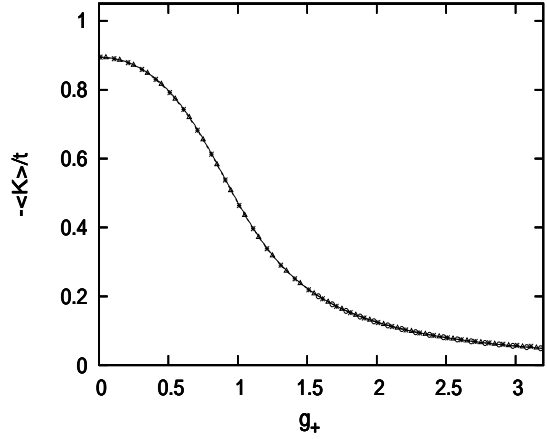}}}
\vspace*{0.5cm}
\caption[]{
Convergence of the kinetic energy $- \langle K \rangle /t$ for the 
two-site system. 
Solid line: $t$=1.0, $\epsilon_d$=1.0 with $n_p$=40,
solid squares are for $n_p$=15, solid triangles are for $n_p$=20,
solid circles are for $n_p$=25.
All energy parameters are expressed in units of $\omega$=1.
}
\label{scaling}
\end{figure}
\begin{figure}
\resizebox*{3.1in}{2.5in}{\rotatebox{270}{\includegraphics{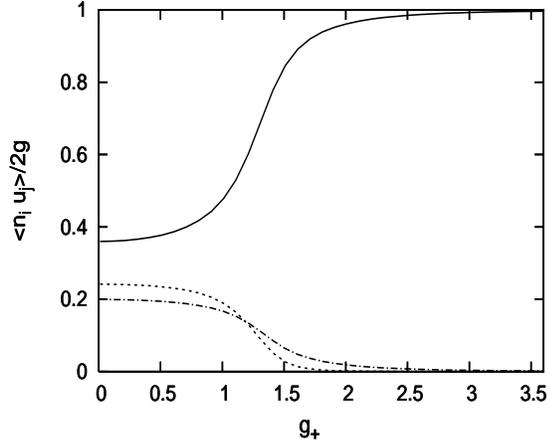}}}
\vspace*{0.5cm}
 \caption[]{
Variation of $\langle n_iu_j \rangle /2g$ with $g_+$ for $t=2.1$ and 
$\epsilon_d$=1.0 for the two-site Holstein model. 
Solid line:$\langle n_1u_1\rangle/2g$, 
dashed line:$\langle n_2u_2\rangle /2g$ and dash-dotted 
line:$\langle n_1u_2 \rangle /2g$=$\langle n_2u_1 \rangle /2g$.
}
\label{scaling}
\end{figure}
\begin{figure}
\resizebox*{3.1in}{2.5in}{\rotatebox{270}{\includegraphics{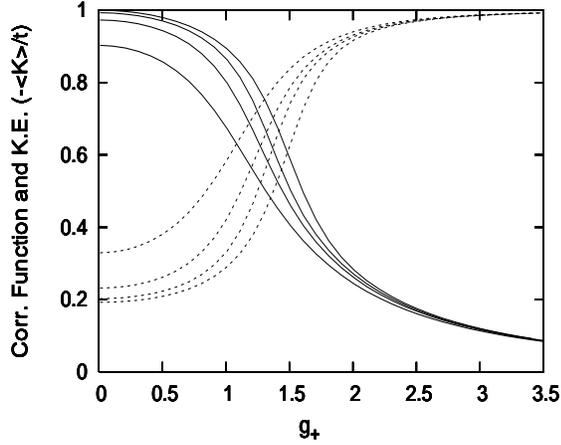}}}
\vspace*{0.5cm}
\caption[]{
Variation of the kinetic energy $- \langle K \rangle /t$ (solid lines) 
and the correlation function $\chi_d^{avg}$ (dashed lines) with $g_+$ 
for t=2.1 and different $\epsilon_d$ for the two-site Holstein model. 
Solid lines from top to bottom and dashed lines from bottom to top 
are for $\epsilon_d$.=0, 0.5, 1.0 and 2.0, respectively.
}
\label{scaling}
\end{figure}
\begin{figure}
\resizebox*{3.1in}{2.5in}{\rotatebox{0}{\includegraphics{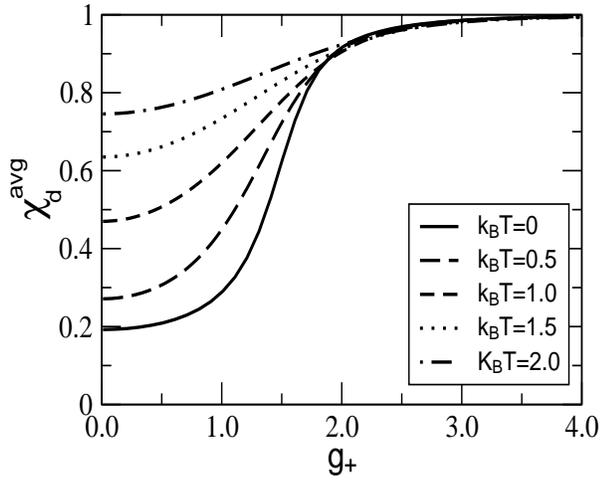}}}
\vspace*{0.5cm}
\caption[]{
Variation of $\chi_d$ with $g_+$ for the ordered case ($\epsilon_d=0$) 
for different temperatures. Solid line: $k_BT=$0, 
long dashed line: $k_BT=$0.5, short dashed line: $k_BT=$1.0, 
dotted line: $k_BT=$1.5, dash-dotted line $k_BT=$2.0. 
$t=2.1$ in the energy scale of $\omega$=1.
}
\label{scaling}
\end{figure}
\vskip 2.5in
\begin{figure}
\resizebox*{3.1in}{2.5in}{\rotatebox{270}{\includegraphics{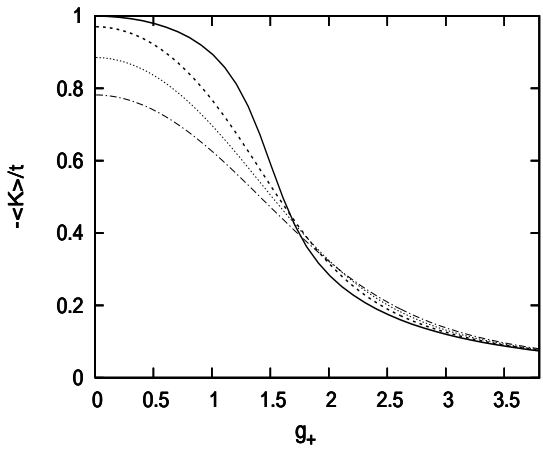}}}
\vspace*{0.5cm}
\caption[]{
Variation of the kinetic energy $-\langle K\rangle /t$ with $g_+$ 
for the ordered 
case for different temperatures. Solid line: $k_BT=$0, 
short dashed line: $k_BT=$1.0, 
dotted line: $k_BT=$1.5, dash-dotted line $k_BT=$2.0.
$t=2.1$ in the energy scale of $\omega$=1.
}
\label{scaling}
\end{figure}
\begin{figure}
\resizebox*{3.1in}{2.5in}{\rotatebox{270}{\includegraphics{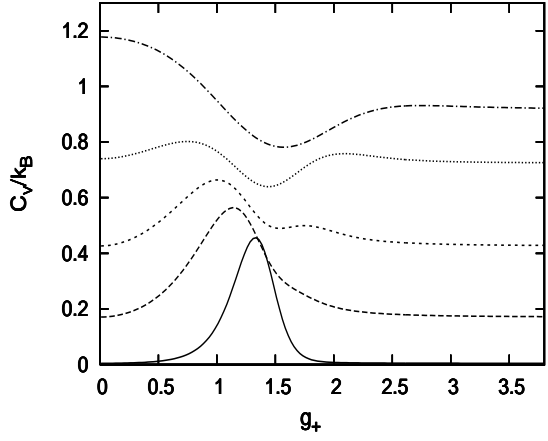}}}
\vspace*{0.5cm}
\caption[]{
Variation of the specific heat ($C_v/k_B$) with $g_+$ for the ordered 
two-site Holstein model for different temperatures. 
Solid line: $k_BT=$0.1, long dashed line: $k_BT=$0.2, 
short dashed line: $k_BT=$0.3, 
dotted line: $k_BT=$0.4, dash-dotted line $k_BT=$0.5.
$t=2.1$ in the energy scale of $\omega$=1.
}
\label{scaling}
\end{figure}
\begin{figure}
\resizebox*{3.1in}{2.5in}{\rotatebox{270}{\includegraphics{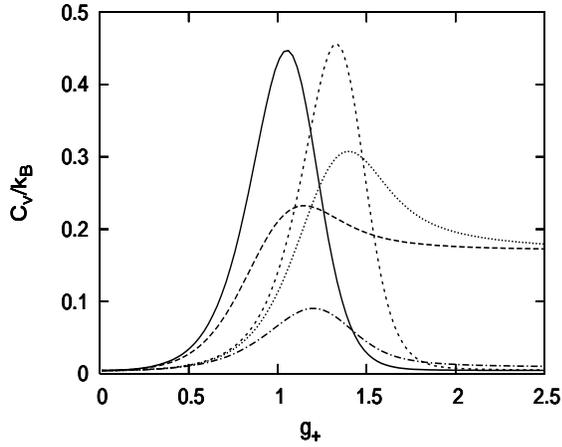}}}
\vspace*{0.5cm}
\caption[]{
Variation of the specific heat ($C_v/k_B$) with $g_+$ at a temperature
$k_BT$=0.1 for ordered and disordered two-site Holstein model 
for different hopping parameters. 
Solid line: $t=$1.0,$\epsilon_d$=0; long dashed line: $t=$1.0,$\epsilon_d$=0.5;
short dashed line: $t=$2.1,$\epsilon_d$=0; 
dotted line: $t=$2.1,$\epsilon_d$=0.5; 
dash-dotted line $t=$2.1,$\epsilon_d$=1.0.
}
\label{scaling}
\end{figure}
\begin{figure}
\resizebox*{3.1in}{2.5in}{\rotatebox{270}{\includegraphics{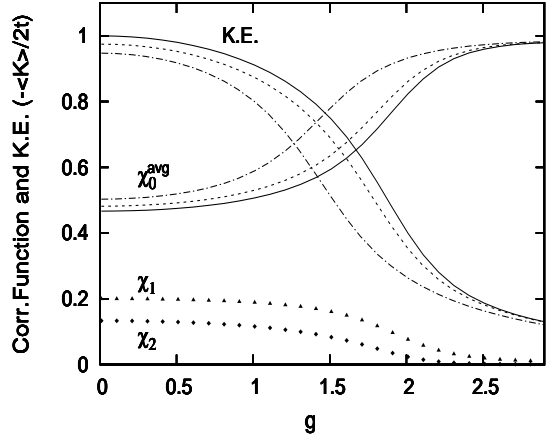}}}
\vspace*{0.5cm}
\caption[]{
Variation of the kinetic energy $- \langle K\rangle /2t$ and the 
correlation function 
$\chi_0^{avg}$ with $g$ in the ground state for ordered and disordered 
four-site Holstein model. 
Solid lines:ordered case, dashed lines:disordered with site potentials
(1,0,0,0), dash-dotted curves: site potentials (-1,0,0,0).
Solid triangles and diamonds represent $\chi_1$ and $\chi_2$, 
respectively for the ordered case. 
Value of the hopping parameter used $t=1.0$ in the energy scale of $\omega$=1.
}
\label{scaling}
\end{figure}
\newpage
\begin{figure}[h]
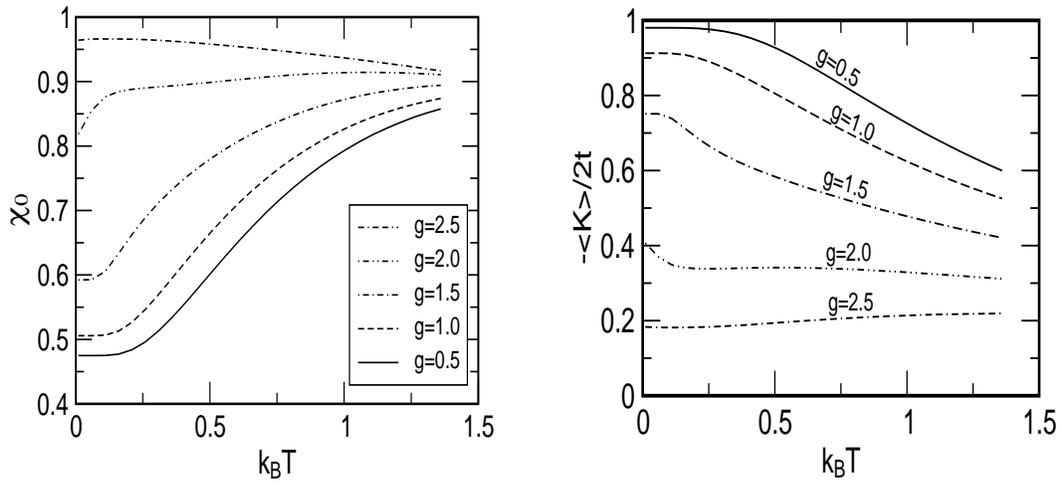

%\begin{tabular}{ll}
%\begin{minipage}{3.0 in}
\resizebox*{5.5in}{2.5in}{\includegraphics[height=2.0in,width=2.5in]{fig10a.ps} 
%\resizebox*{5.1in}{5.0in}{\rotatebox{0}{\includegraphics{fig10.ps}}}
%\end{minipage} 
%&
%\begin{minipage}{3.0in}
\hspace{.3in}
\includegraphics[height=2.0in,width=2.5in]{fig10b.ps}} 
%\end{minipage}
%\end{tabular}
\vspace*{0.5cm}
\caption[]{
Variation of $\chi_0$ and the kinetic energy $- \langle K \rangle/2t$
with temperature for the ordered four-site Holstein model for different 
values of $g$. Value of the hopping parameter used $t=1.0$ in the 
energy scale of $\omega$=1.
}
\label{scaling}
\end{figure}
\newpage
\begin{figure}
\resizebox*{3.1in}{2.5in}{\rotatebox{270}{\includegraphics{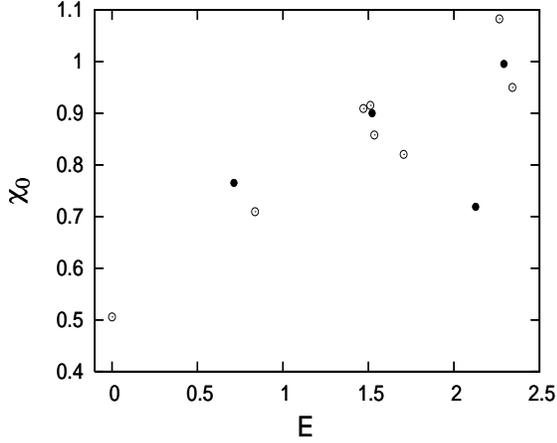}}}
\vspace*{0.5cm}
\caption[]{
Values of the on-site correlation function ($\chi_0$) for the ground
and the excited states for $g=1$ and $t=1$ for the ordered four-site 
Holstein model shown up to $E=2.4$, where $E$ is the energy of the excited 
state measured from the ground state energy. There are sixteen states
within the energy range $0\le E\le 2.4$. Four states have double 
degeneracy (shown by solid circles).
}
\label{scaling}
\end{figure}
\begin{figure}
\resizebox*{3.1in}{2.5in}{\rotatebox{270}{\includegraphics{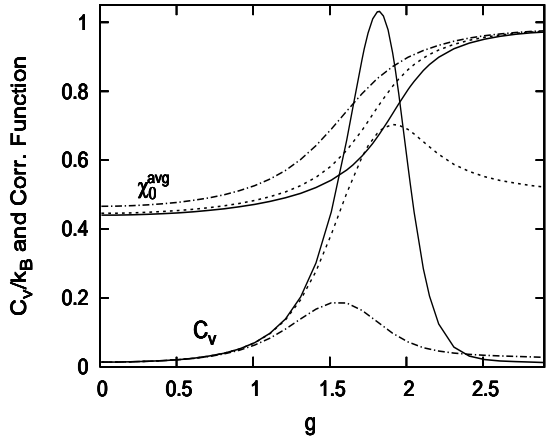}}}
\vspace*{0.5cm}
\caption[]{
Variation of the specific heat ($C_v/k_B$) and the correlation 
functions $\chi_0^{avg}$ with $g$ at a temperature
$k_BT$=0.1 for $t$=1.2 for different set of site potentials for 
four-site Holstein model. 
Solid lines:ordered case, dashed line: disorderd case with site
potentials (-0.5,0,0,0) and dash-dotted line: disordered case with site
potentials (-1,0,0,0). 
}
\label{scaling}
\end{figure}
\newpage
\begin{figure}
\resizebox*{3.1in}{2.5in}{\rotatebox{270}{\includegraphics{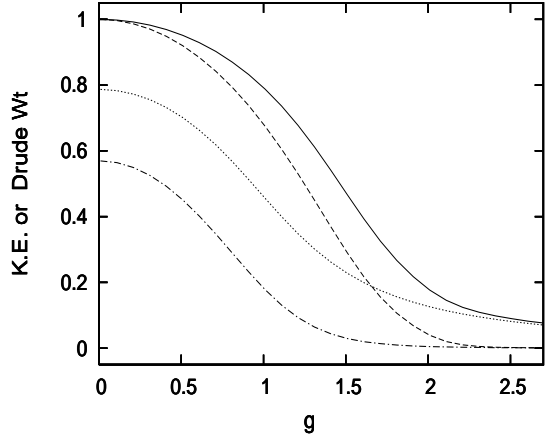}}}
\vspace*{0.5cm}
\caption[]{
Variation of the kinetic energy  $- \langle K\rangle /2t$ and the 
Drude weight 
with $g$ in the ground state for ordered and disordered cases
for $t$=0.5.  
Ordered case: Kinetic energy (solid line), Drude weight (dashed line).
Disordered case with site potentials (-1,0,0,0): 
Kinetic energy (dotted line), Drude weight (dash-dotted line).
}
\label{scaling}
\end{figure}
\begin{figure}
\resizebox*{5.1in}{4.0in}{\rotatebox{270}{\includegraphics{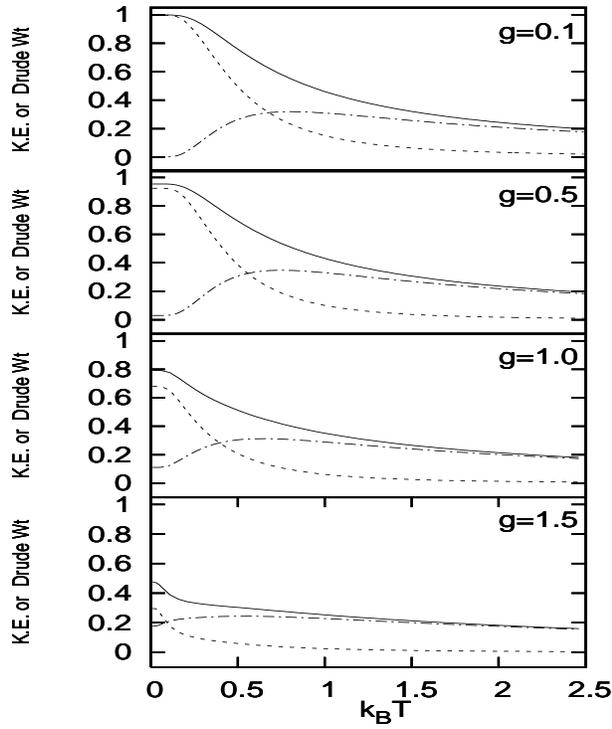}}}
\vspace*{0.5cm}
\caption[]{
Variation of the kinetic energy $- \langle K \rangle /2t$ (solid lines) 
and the Drude weight ($D/2t$) (dashed lines) with $k_BT$ for the 
ordered lattice 
for different values of $g$. The dot-dashed lines represent the 
incoherent part of the kinetic energy. 
Value of the hopping parameter used $t$=0.5.  
}
\label{scaling}
\end{figure}
\newpage
\begin{figure}
\resizebox*{3.1in}{2.5in}{\rotatebox{270}{\includegraphics{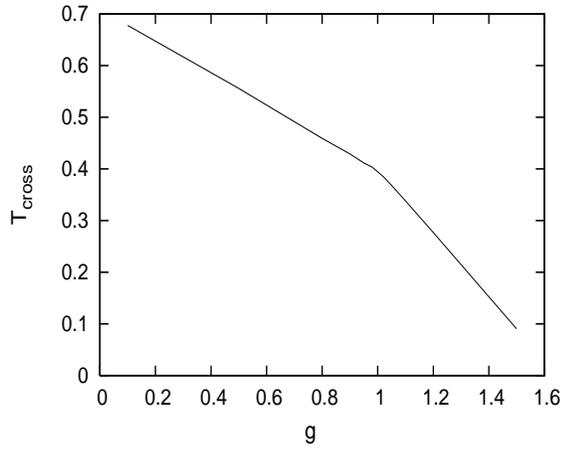}}}
\vspace*{0.5cm}
\caption[]{
Variation of the cross-over temperature ($T_{cross}$) with $g$ for 
$t$=0.5 for the ordered case.
}
\label{scaling}
\end{figure}
\begin{figure}
\resizebox*{3.1in}{2.5in}{\rotatebox{270}{\includegraphics{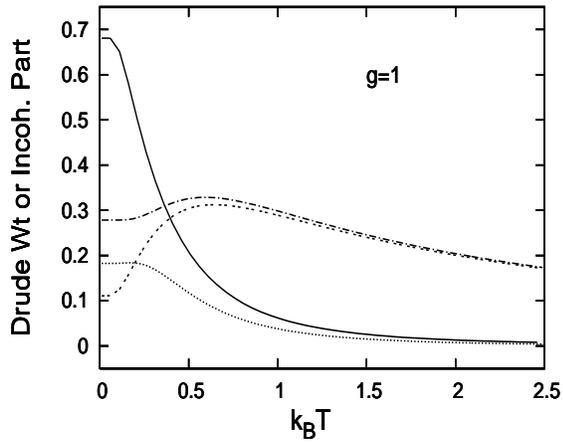}}}
\vspace*{0.5cm}
\caption[]{
Variation of the Drude weight and the incoherent part of the kinetic
energy (in a scale of 2$t$) with temperature for ordered and disordered 
cases for $t$=0.5.
Ordered case: Drude weight (solid line), incoherent part of the
K.E. (dashed line).
Disordered case with site potentials (-1,0,0,0):
Drude weight (dotted line), incoherent part of the K.E. (dash-dotted line).
}
\label{scaling}
\end{figure}
\end{document}